\documentclass[pdflatex,sn-nature,Numbered]{sn-jnl}
\usepackage{graphicx}
\usepackage{multirow}
\usepackage{amsmath,amssymb,amsfonts}
\usepackage{amsthm}
\usepackage{mathrsfs}
\usepackage{xcolor}
\usepackage{csquotes}
\usepackage{textcomp}
\usepackage{manyfoot}
\usepackage{booktabs}
\usepackage{listings}
\newcommand{\appropto}{\mathrel{\vcenter{
  \offinterlineskip\halign{\hfil$##$\cr
    \propto\cr\noalign{\kern2pt}\sim\cr\noalign{\kern-2pt}}}}}

\begin{document}

\title[Crosstalk effects in microwave SQUID multiplexed TES bolometer readout]{Crosstalk effects in microwave SQUID multiplexed TES bolometer readout}

%%%%%%%%%%%%%%%%%%%%%%%%%%%%
% Authors and institutions %
%%%%%%%%%%%%%%%%%%%%%%%%%%%%
\author*[1,2]{\fnm{John C.} \sur{Groh}}\email{john.groh@lbl.gov}
\author[3,4]{\fnm{Zeeshan} \sur{Ahmed}}
\author[3,4]{\fnm{Shawn W.} \sur{Henderson}}
\author[1]{\fnm{Johannes} \sur{Hubmayr}}
\author[1]{\fnm{John A. B.} \sur{Mates}}
\author[5,6]{\fnm{Maximiliano} \sur{Silva-Feaver}}
\author[1,7]{\fnm{Joel} \sur{Ullom}}
\author[4,8]{\fnm{Cyndia} \sur{Yu}}

\affil[1]{\orgdiv{Quantum Sensors Division}, \orgname{National Institute of Standards and Technology}, \orgaddress{\city{Boulder}, \postcode{80305}, \state{CO}, \country{USA}}}
\affil[2]{\orgdiv{Physics Division}, \orgname{Lawrence Berkeley National Laboratory}, \orgaddress{\city{Berkeley}, \postcode{94720}, \state{CA}, \country{USA}}}
\affil[3]{\orgname{Kavli Institute of Particle Astrophysics and Cosmology}, \orgaddress{\city{Menlo Park}, \postcode{94025}, \state{CA}, \country{USA}}}
\affil[4]{\orgname{SLAC National Accelerator Laboratory}, \orgaddress{\city{Menlo Park}, \postcode{94025}, \state{CA}, \country{USA}}}
\affil[5]{\orgdiv{Department of Physics}, \orgname{Yale University}, \orgaddress{\city{New Haven}, \postcode{06520}, \state{CT}, \country{USA}}}
\affil[6]{\orgdiv{Wright Laboratory, Department of Physics}, \orgname{Yale University}, \orgaddress{\city{New Haven}, \postcode{06520}, \state{CT}, \country{USA}}}
\affil[7]{\orgdiv{Department of Physics}, \orgname{University of Colorado Boulder}, \orgaddress{\city{Boulder}, \postcode{80309}, \state{CO}, \country{USA}}}
\affil[8]{\orgdiv{Department of Physics}, \orgname{Stanford University}, \orgaddress{\city{Stanford}, \postcode{94305}, \state{CA}, \country{USA}}}

%%%%%%%%%%%%
% Abstract %
%%%%%%%%%%%%
\abstract{

% TES-based detectors super useful, many applications which require scaling # of detectors
% umux offers a scaling solution 
Transition-edge sensor (TES) bolometers are broadly used for background-limited astrophysical measurements from the far-infrared to mm-waves.  Many planned future instruments require increasingly large detector arrays, but their scalability is limited by their cryogenic readout electronics.  Microwave SQUID multiplexing offers a highly capable scaling solution through the use of inherently broadband circuitry, enabling readout of hundreds to thousands of channels per microwave line.  As with any multiplexing technique, the channelization mechanism gives rise to electrical crosstalk which must be understood and controlled so as to not degrade the instrument sensitivity.  Here, we explore implications relevant for TES bolometer array applications, focusing in particular on upcoming mm-wave observatories such as the Simons Observatory and AliCPT.  We model the relative contributions of the various underlying crosstalk mechanisms, evaluate the difference between fixed-tone and tone-tracking readout systems, and discuss ways in which crosstalk nonlinearity will complicate on-sky measurements.}

\keywords{crosstalk, multiplexing, microwave SQUID, TES}

\maketitle

%%%%%%%%%%%%%%%%
% Introduction %
%%%%%%%%%%%%%%%%
\section{Introduction}

\par
TES bolometers remain one of the most commonly used detectors for background-limited astrophysical imaging observations from mm wavelengths to the far-infrared \cite{henderson2016advactpol, stebor2016simonsarray, harrington2016class, bergman2018spider2, sobrin2022spt3g, ade2022bicep3, holland2013scuba2, harper2018hawcplus, zhang2020aste}, with several proposed and planned future observatories requiring more than $10^5$ sensors \cite{ade2019sosciencegoals, abazajian2019cmbs4, sehgal2019cmbhd}.  Such large detector counts necessitate highly multiplexed readout to manage system complexity, control thermal loading on the sub-Kelvin cooling system, and reduce the total implementation cost.  Of the several cryogenic multiplexing schemes developed for TES bolometer readout, microwave SQUID multiplexing has demonstrated the highest channel carrying capacity, with a 910-channel readout module recently developed for cosmic microwave background (CMB) imaging applications \cite{mccarrick2021ufm}.

\par
\begin{figure}[htbp]
\includegraphics[width=\textwidth]{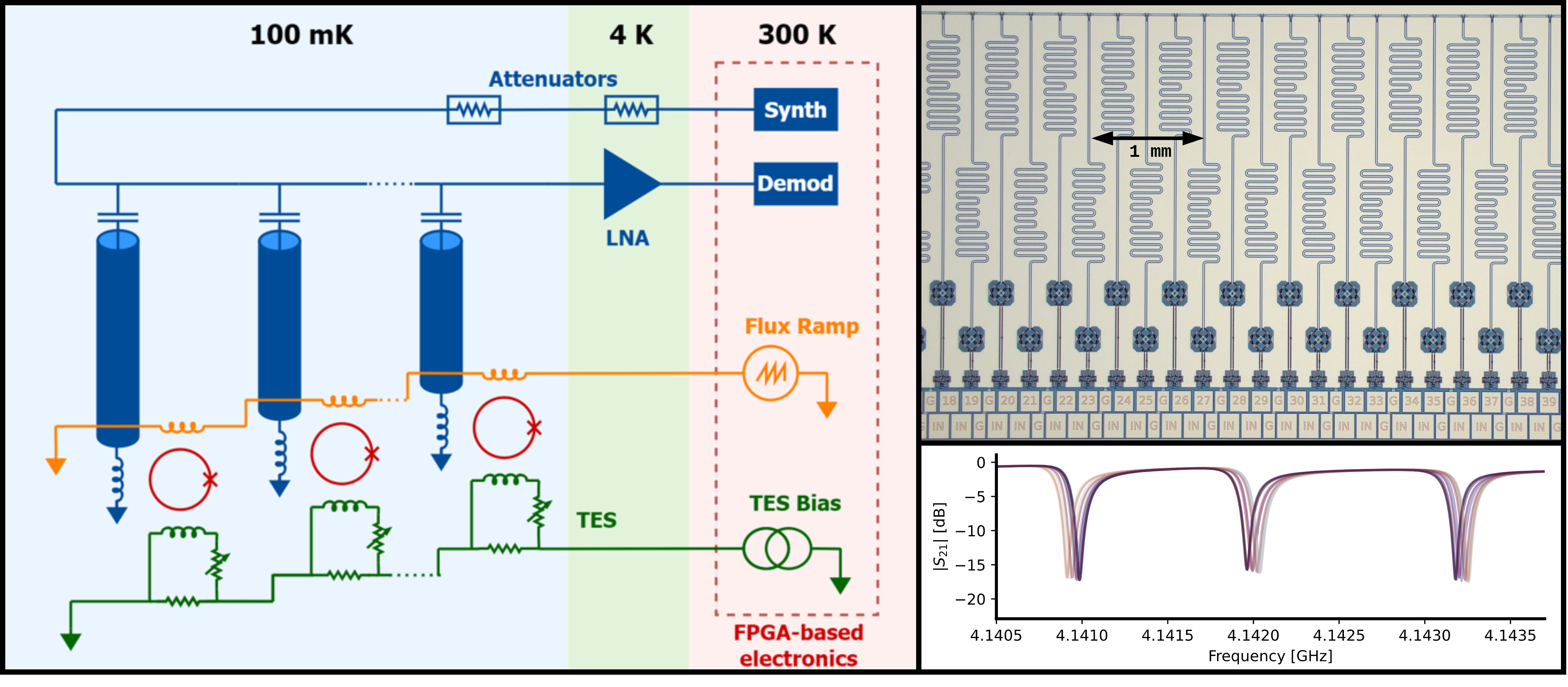}
\caption{(Color figure online) \textit{Left:} Circuit schematic of a microwave SQUID multiplexer.  \textit{Top right:} Micrograph of a chip containing the sub-Kelvin multiplexer components, with bond pads at the bottom for hybridization with a TES array.  \textit{Bottom right:}  Measurement of multiple SQUID-coupled resonators on the same readout line simultaneously being modulated by applied magnetic flux (color indicates the applied flux, which spans a range of 1 magnetic flux quantum). \label{fig:principle}}
\end{figure}

In a microwave SQUID multiplexer, channelization is achieved by coupling each detector to a unique microwave resonator (see Figure~\ref{fig:principle}).  With modern photolithography techniques, hundreds of superconducting resonators with $Q \sim 5\times 10^4$ may now be reliably fabricated on a few square centimeters of silicon with unique and regularly spaced resonance frequencies in the $\sim$5 GHz range.  These resonators are capacitively coupled to a common microwave feedline, through which interrogation tones continuously probe the changing resonance frequencies as they are modulated by the detector signals.  This modulation is achieved through an inductively coupled rf-SQUID, which effectively tunes part of the resonator inductance.  An additional layer of \enquote{flux ramp} modulation is applied to all channels, linearizing the SQUIDs and mitigating the effect of low frequency noise from two-level system fluctuations \cite{mates2012fluxramp}.  This modulation ensures that the signal in the $i^{\mathrm{th}}$ detector is directly proportional to the phase $\phi_i$ of the modulated resonance frequency $f_i$.

\par
Electrical crosstalk is a complication with which all readout schemes must contend, though the physical mechanisms responsible for the phenomena vary from scheme to scheme.  With microwave SQUID multiplexed readout, the physics behind and effects of crosstalk are somewhat more complex than in competing multiplexing methods, meriting further study.  The impact of crosstalk is also highly application dependent.  In these proceedings, we focus on bolometric imaging of the CMB, for which electrical crosstalk is typically targeted to be $<$1\%.  Two prior instruments - MUSTANG-2 on the GBT \cite{dicker2014mustang} and a retrofitted Keck Array reciever \cite{cukierman2020keck,yuphdthesis} - have demonstrated initial technical feasibility of microwave SQUID multiplexing with TES bolometers on the sky, but each with relatively small ($\mathcal{O}(100)$) detector counts.  At least two upcoming CMB instruments -- the Simons Observatory~\cite{ade2019sosciencegoals} and AliCPT~\cite{salatino2020alicptreceiver} -- will soon begin deep surveys of the mm-wave sky using microwave SQUID multiplexing readout of $>10^4$ TES bolometers each.  Understanding the nature and magnitude of multiplexer-induced crosstalk is required in order to achieve the science goals of these observatories.

%\par
%These proceedings are structured as follows.  In Section~\ref{sec:mechanisms}, we summarize the known physical mechanisms which contribute to the net electrical crosstalk and model their magnitudes for upcoming CMB applications.  In Section~\ref{sec:tracking}, we evaluate the impact of tone-tracking readout, a new development in which the resonator interrogation tones follow the resonance minima in time to minimize the transmitted power through the cryogenic amplification chain.  We then discuss the nonlinear nature of the dominant crosstalk mechanism and its impact on CMB measurements in Section~\ref{sec:nonlinear} before concluding in Section~\ref{sec:conclusion}.

%%%%%%%%%%%%%%%%%%%%%%%%%%%
% Contributing mechanisms %
%%%%%%%%%%%%%%%%%%%%%%%%%%%
\section{Contributing mechanisms}
\label{sec:mechanisms}
\par
Prior work has elucidated four physical mechanisms which introduce crosstalk in microwave SQUID multiplexed readout \cite{mates2019umuxcrosstalk}.  %The four mechanisms are due to 1.) pairwise hybridization of resonators as electrical oscillators, 2.) parasitic inductive coupling between inductive elements involved in the detector-to-resonator coupling circuit, 3.) off-resonance transmission of neighboring resonances on the same feedline, and 4.) 3-wave mixing products arising from system nonlinearities.
Sections~\ref{subsec:CHO}--\ref{subsec:3wave} address these effects for the specific case of TES bolometer readout.  In Section~\ref{subsec:estimate} we provide an estimate for the magnitude of all these mechanisms in recent microwave SQUID multiplexer chips optimized for TES bolometer readout\footnote{Specifically, the results in this paper apply to the NIST $\mu$mux100k v3.3.2 design, which was used for the Simons Observatory multiplexer production and for the first light detector module of AliCPT.} \cite{dober2021muxchip}.

%%% CHO subsection %%%
\subsection{Harmonic oscillator coupling}
\label{subsec:CHO}

\par
\begin{figure}[htbp]
    \centering
    \includegraphics[width=0.39\linewidth]{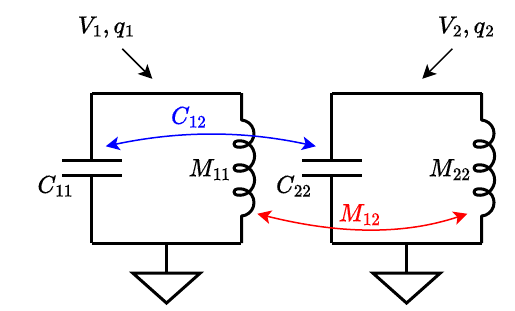}
    \includegraphics[width=0.59\linewidth]{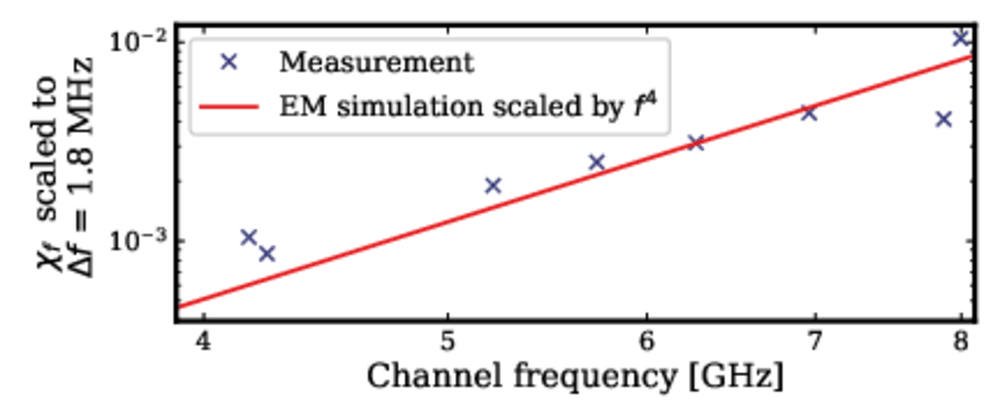}
    \caption{(Color figure online) \textit{Left:} Simplified circuit model for coupled oscillator crosstalk.  \textit{Right:} Measurement and extrapolated simulation of coupled oscillator crosstalk over the 4-8 GHz readout octave used to confirm the  $\bar{f}^4$ dependence of $\chi_f$ from Equation~\ref{eq:cho}.}
    \label{fig:cho}
\end{figure}

The dominant mechanism of crosstalk for upcoming bolometric applications is due to pairwise hybridization of resonators; each resonator is imperfectly isolated from its neighbors, so shifts in the resonance position of its neighbors modify its own resonance frequency.  A conceptually helpful model for this crosstalk may be formed using the equivalence of quarter-wave transmission line resonators with characteristic impedances $Z_0$ and resonance frequencies $\omega_i$ to LC resonators with self-inductances $M_{ii} = \frac{4Z_0}{\pi\omega_i}$ and self-capacitances $C_{ii} = \frac{\pi}{4 Z_0\omega_i}$, as shown on the left in Figure~\ref{fig:cho}.

\par
If there is a parasitic mutual inductive coupling $M_{12}$ and a parasitic mutual capacitive coupling $C_{12}$, the voltages $V_i$ will be related to the circulating currents $I_i$ and accumulated charges $q_i$ via
\begin{equation}
V_i = \sum_j M_{ij}I_j = \sum_j P_{ij}q_j
\end{equation}
where $\bf{P} = \bf{C}^{-1}$.  In terms of the uncoupled resonance frequencies $\omega_i \equiv \sqrt{P_{ii} / M_{ii}}$ and the coupling parameters $\alpha\equiv M_{12} / M_{11}\approx M_{12} / M_{22}$ and $\omega_p\equiv\sqrt{P_{12}/M_{11}}\approx\sqrt{P_{12}/M_{22}}$, nontrivial oscillating solutions to Kirchoff's equations for the charges have frequencies
\begin{equation}
    \begin{array}{lr}
\omega_1'=\sqrt{\omega_1^2 + \frac{2\omega_1^2\omega_2^2\alpha^2 - (\omega_1^2+\omega_2^2)\omega_p^2\alpha + 2\omega_p^4}{\omega_1^2 - \omega_2^2}}\\
\omega_2'=\sqrt{\omega_2^2 - \frac{2\omega_1^2\omega_2^2\alpha^2 - (\omega_2^2-\omega_2^2)\omega_p^2\alpha + 2\omega_p^4}{\omega_1^2 - \omega_2^2}}
    \end{array}
\end{equation}
to lowest order in the small parameters $\alpha$ and $\frac{\omega_p^2}{\omega_1^2 - \omega_2^2}$.  The frequencies $\omega_1'$ and $\omega_2'$ represent the modified self-resonances of the two channels.  In the limit where $\bar{\omega}\approx\omega_1\approx\omega_2\approx\omega_2'$, these hybridized eigenmodes of the 2-oscillator system introduce the fractional frequency crosstalk
\begin{equation}
\label{eq:cho}
    \chi_f\equiv\frac{df_2'}{df_1} = \frac{\bar{f}^4}{2(\Delta f)^2}\left(\frac{\pi^2}{16Z_0^2}M_{12}^2 + M_{12}C_{12} + \frac{16Z_0^2}{\pi^2}C_{12}^2\right) = \gamma\frac{\bar{f}^4}{(\Delta f)^2}.
\end{equation}
Practically, the relevant coupling strengths $M_{12}$ and $C_{12}$ are reduced by distributing channels such that spatial neighbors are far apart in frequency space and vice versa \cite{mates2019umuxcrosstalk}.  In the design studied here, crosstalk is largest between nearest frequency neighbors, which are spaced four physical channels apart \cite{dober2021muxchip}.

\par
While in reality the oscillators are distributed structures without lumped parasitic couplings $M_{12}$ and $C_{12}$, this result demonstrates the strong dependence on channel frequency $\bar{f}$ and pairwise channel spacing $\Delta f$.  An additional complication is that the geometry of quarter-wave resonators changes significantly over the octave of bandwidth in which they resonate, so the prefactor $\gamma$ is potentially itself a function of $\bar{f}$.  To address these complications, we performed a full electromagnetic simulation of ten adjacent resonators with AWR Microwave Office\footnote{\url{https://www.cadence.com/en_US/home/tools/system-analysis/rf-microwave-design/awr-microwave-office.html}} to confirm the $(\Delta f)^{-2}$ scaling and direct measurements (shown on the right in Fig.~\ref{fig:cho}) of real devices to determine that the absolute frequency scaling is remarkably close to the modeled $\bar{f}^4$ scaling with a best fit $\gamma$ of $9.6\times 10^{-30}\;\mathrm{Hz}^{-2}$.

\par
In addition to the modeling results from Section~\ref{subsec:estimate}, we confirm this resonator hybridization is the dominant crosstalk mechanism through several observations.  First, we note that the direct measurements of overall crosstalk shown on the right in Fig.~\ref{fig:cho} find a positive crosstalk with a strong dependence on the absolute frequency $\bar{f}$; these distinguish it from Lorentzian crosstalk. 
 Additionally, we observe that it couples resonance frequencies, as opposed to the phase of the flux ramp modulation, distinguishing it from direct inductive crosstalk or nonlinear mixing crosstalk.

%%% Direct inductive subsection %%%
\subsection{Direct inductive coupling}

\par
A second mechanism which induces crosstalk is due to parasitic mutual inductances between the explicit coupling inductors in physically adjacent readout channels; a signal current in one channel directly induces crosstalk currents in its neighbors through through these inter-channel transformer couplings.  The inductor which transduces a perpetrator channel's TES current into a magnetic flux in its SQUID can also induce a magnetic flux in a neighboring victim SQUID.  Additionally, the screening current in a perpetrator channel's SQUID also induces a magnetic flux in a neighboring victim SQUID.  These couplings induce a fractional phase crosstalk according to
\begin{equation}
    \chi_\phi \equiv \cfrac{d\phi'_2}{d\phi_1} = 
    \begin{cases}
        \cfrac{M_{i1,s2}}{M_{i1,s1}}\;\;\; \mathrm{in\; the\; case\; of\; parasitic\; input-SQUID\; coupling}\\
        \cfrac{M_{s1,s2}}{M_{s1,s1}}\;\;\; \mathrm{in\; the\; case\; of\; parasitic\; SQUID-SQUID\; coupling}
    \end{cases}
\end{equation}
where the subscript $i1$ refers to the input inductor of channel 1, the subscript $s2$ refers to the SQUID loop of channel 2, etc.  We extract the relevant inductances from FastHenry\footnote{\url{https://www.fastfieldsolvers.com/fasthenry2.htm}} simulations of the multiplexer chip layout (shown in Table~\ref{tab:couplings}).  Cross-channel mutual inductances are strongly suppressed by the second order gradiometric layout of the SQUID cells.\\

% direct inductive couplings table
% something with the formatting is still wrong, but I haven't figured it out yet...
\begin{table}[h]
  \centering
  \caption{Simulated self inductances and mutual inductances between physically adjacent channels which are relevant for direct inductive coupling crosstalk.}
  \label{tab:couplings}
  %\centering
  \begin{tabular}{| c | c | c | c |}
    \hline      $M_{i1,s1}$ & $M_{s1,s1}$ & $M_{i1,s2}$ & $M_{s1,s2}$ \\
    \hline      227 pH & 22.4 pH & 32 fH & 2 fH\\
    \hline
  \end{tabular}
\end{table}

%%% Lorentzian subsection %%%
\subsection{Lorentzian combination}

\par
Because each resonator is coupled to the same feedline, the off-resonance transmission of one resonator slightly affects the transmission of its neighbors and gives rise to crosstalk.  Quantitatively, the combined transmission $S_{31}$ of two resonators separated by a distance $\ell$, the first with S-matrix $S$ connecting ports 1 and 2 and the second with S-matrix $S'$ connecting ports 2 and 3, coupled to a common feedline with phase velocity $v_p$ is
\begin{equation}
    S_{31} = \frac{S_{21}S'_{32}}{1 - S_{11}S'_{22}e^{-i\omega\ell/v_p}}
\end{equation}
 up to an overall phase factor which will be removed by the readout \cite{mates2019umuxcrosstalk}.  For the tone-tracking readout in use for upcoming bolometric applications, the tracking feedback enforces that Im$(S_{31})$ vanishes \cite{yu2023smurf}.  Given a simulated detector signal in a perpetrator channel, we therefore numerically solve for the time-dependent frequency which satisfies the tone tracking feedback condition to generate a crosstalk signal in a victim channel, inputting design values for the phase delay between the two resonators along the shared feedline.  In general there is no simple analytic expression for the magnitude of this form of crosstalk, but for reasonable channel parameters we find that $\chi_f \appropto -1/n^2$ where $n$ is the number of resonator bandwidths by which the two channels are spaced in frequency.

\subsection{Nonlinear mixing}
\label{subsec:3wave}

\par
\begin{figure}[htbp]
\begin{center}
\includegraphics[width=0.8\textwidth]{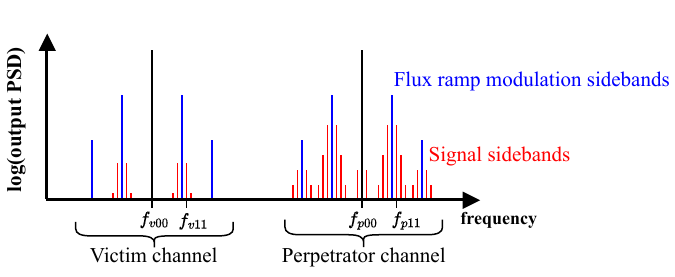}
\caption{Illustration of how mixing of two large-amplitude features such as those at the resonance frequencies $f_{v00}$ and $f_{p00}$ of a victim and perpetrator channel, along with a small-amplitude signal sideband $f_{p11}$ of the perpetrator channel can create a signal-like sideband in the victim channel at $f_{v11} = f_{v00} - f_{p00} + f_{p11}$. \label{fig:mixing}}
\end{center}
\end{figure}

A final mechanism of crosstalk arises when intermodulation products caused by nonlinearity in the cryogenic amplification chain mimic the spectral structure of a real signal at the input to the demodulator.  The RF spectrum of a flux ramp modulated signal with frequency $f_s$ after a stage of nonlinear amplification contains sidebands at locations $f_{ijk} = f_i + jf_m + kf_s$, where $f_i$ is the $i^{\mathrm{th}}$ resonator frequency, $f_m$ is the flux ramp modulation rate, and $j$ and $k$ are integers \cite{yu2022bandwidthaliasing}.  Nonlinear mixing of three tones $\{f_{ijk},\, f_{i'j'k'},\, f_{ij''k''}\}$ with $k = k' = 0$, $k'' \neq 0$, and $i \neq i'$ will produce a signal-like sideband in a victim channel as illustrated in Figure~\ref{fig:mixing}.  This happens simultaneously for all $k \neq 0$ sidebands, producing a copy of the full phase modulated signal spectrum in the victim channel and inducing a fractional phase crosstalk of
\begin{equation}
\label{eq:IMD}
\chi_\phi \approx \frac{4P_{tone}}{P_{IIP3}}
\end{equation}
where $P_{tone}$ is the power in the largest RF feature transmitted to the amplifier with $k = 0$ and $P_{IIP3}$ is the input third-order intercept point of the amplifier chain \cite{mates2019umuxcrosstalk}.  As this mechanism copies the phase modulation sidebands from one channel onto another, it mixes flux ramp demodulation phases rather than resonance frequencies.  In reality, the exact amplitudes and phases of the features in the transmitted RF spectra are complicated by tone-tracking readout (further discussed in Section~\ref{sec:tracking}), making them difficult to model precisely.  For the purposes of generating a rough estimate of the magnitude of this crosstalk in our devices, we use the simplistic assumption that $P_{tone} < P_{probe}Q^2/Q_i^2$, where $P_{probe}$ is the interrogation tone power on the feedline just before the resonator, together with Equation~\ref{eq:IMD}.  

%\par
%{\color{red}
%This nonlinear mixing crosstalk is not to be confused with a separate mixing phenomenon which is common to all microwave resonator-based multiplexing schemes: namely, the mixing of three RF features with $k = 0$ associated with three unique channels.  Such mixing will not reproduce the spectral shape of a perpetrator signal in a victim channel, so we do not consider it as a mechanism of crosstalk.  Rather, since for modern bolometric instruments the multiplexing factor $N_{\mathrm{mux}}$ is of $\mathcal{O}(10^3)$ and the number of these mixing products grows at least as fast as $N_{\mathrm{mux}}^3$, they together form a broadband pseudo-noise floor.  We leave detailed studies of this phenomenon to future work.}

% nonlinear mixing figure

%\par
%{\color{red}Cut this paragraph?} {\color{blue}It should be noted that this mechanism of crosstalk involves intermodulation products of 2 tones associated with one channel and 1 tone associated with another.  Such products land exactly at frequencies associated with signal sidebands in one of the channels.  In general, 3-wave mixing products of tones associated with three unique channels will land at less controlled frequencies with larger amplitudes.  As the number of these products grows with the cube of the number of input tones, for bolometer readout systems with $\mathcal{O}(10^3)$ channels they present an effective RF floor that acts as a noise source rather than a crosstalk source.  While beyond the scope of this paper, this effective noise floor is an important consideration when designing high channel count systems.}

\subsection{Expected magnitudes in current instruments}
\label{subsec:estimate}

\par
\begin{figure}[htbp]
\begin{center}
\includegraphics[width=\textwidth]{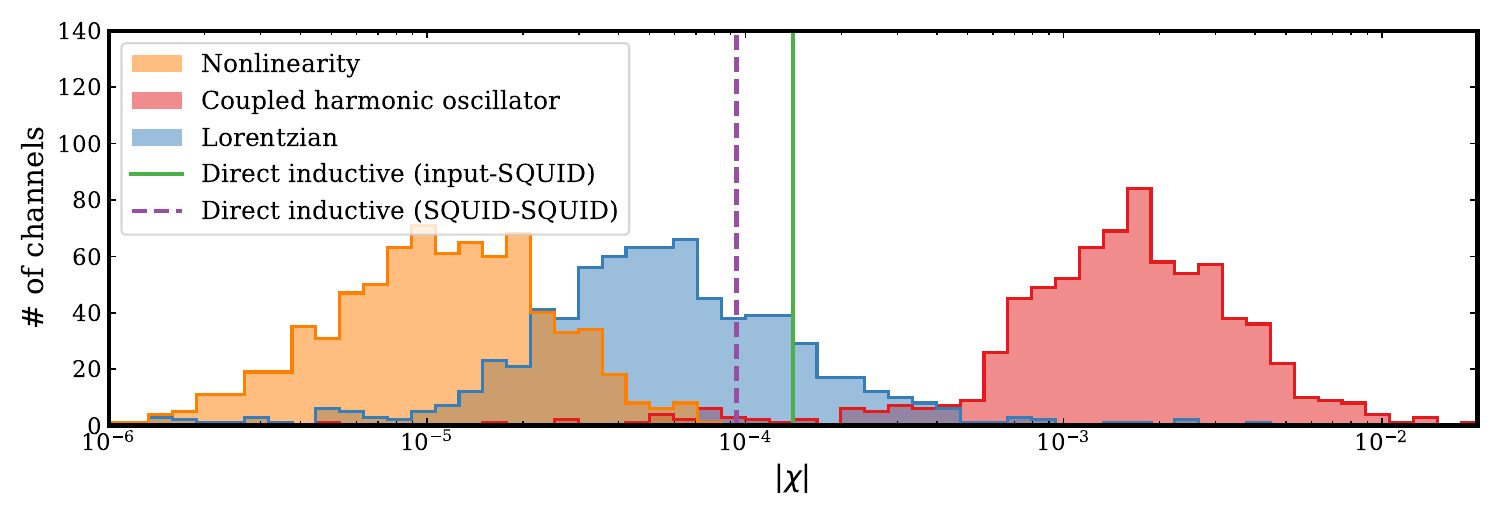}
\caption{(Color figure online) Modeled amplitudes for each mechanism of crosstalk in the microwave SQUID multiplexers planned for use in multiple upcoming mm-wave observatories.  For phase crosstalk mechanisms, binned values are $\lvert\chi_\phi\rvert$ for each channel.  For frequency crosstalk mechanisms, binned values are $\lvert\chi_f\rvert$ for each channel. \label{fig:so_estimate}}
\end{center}
\end{figure}

Using the models described in Sections~\ref{subsec:CHO}--\ref{subsec:3wave}, we model the magnitudes of the four known crosstalk mechanisms for upcoming CMB observatories.  We take as model inputs the measured channel parameters which typically vary due to fabrication processes (resonance frequencies, pairwise frequency spacings, internal and coupling quality factors, and SQUID frequency response amplitudes) from a set of NIST v3.3.2 $\mu$mux100k multiplexer chips \cite{dober2021muxchip}.  For the nonlinear mixing contribution we also assume a probe tone power (-75 dBm) and 3rd-order intercept point (-35 dBm) corresponding to a typical system with a 2-stage cryogenic amplification chain.  As shown in Figure~\ref{fig:so_estimate}, harmonic oscillator coupling is expected to be the dominant mechanism with a typical amplitude of $\chi_f\sim 1.5\times 10^{-3}$, comparable with the amplitude of crosstalk in other multiplexing systems recently used for CMB observations \cite{ade2014bicep2experiment, sobrin2022spt3g}.  Applications desiring reduced crosstalk may consider doing so by increasing the channel spacing at a cost of multiplexing factor, reducing the electromagnetic coupling strength with an alternative resonator layout, further interleaving resonators so that frequency neighbors are even further apart on the chip, and/or improving fabrication tolerances on resonance frequency placement.

%%%%%%%%%%%%%%%%%%%%%%%%%%%%%%%%%%%
% Impact of tone tracking readout %
%%%%%%%%%%%%%%%%%%%%%%%%%%%%%%%%%%%
\section{Impact of tone tracking readout}
\label{sec:tracking}

\par
Upcoming CMB experiments with microwave SQUID multiplexed readout are planning to use tone tracking readout, a new development in which the resonator interrogation tones follow the resonance minima in time to reduce the transmitted power through the cryogenic amplification chain \cite{yu2023smurf}.  In this section, we evaluate the impact of tone tracking schemes on crosstalk.

%\par
%Because microwave SQUID multiplexed readout systems for small-bandwidth detectors such as TES bolometers are able to operate with $N_{channels} \sim 10^3$, 3-wave mixing of resonator probe tones will produce 
%$N_{channels}^3\sim10^9$ intermodulation products at uncontrolled frequencies (to be distinguished from the 3-wave mixing of 2 probe tones and one signal sideband discussed in Section~\ref{subsec:3wave}) which produce an effective noise floor.  As such, it is desirable to implement tone tracking, in which the probe tones actively follow the resonators as they move in frequency to minimize the microwave power incident on the cryogenic amplification chain \cite{yu2023smurf}.  Though tone tracking electronics are primarily motivated by this noise consideration, in this section we evaluate the impact of tone tracking readout schemes on crosstalk.

\par
\begin{figure}[htbp]
\begin{center}
\includegraphics[width=\textwidth]{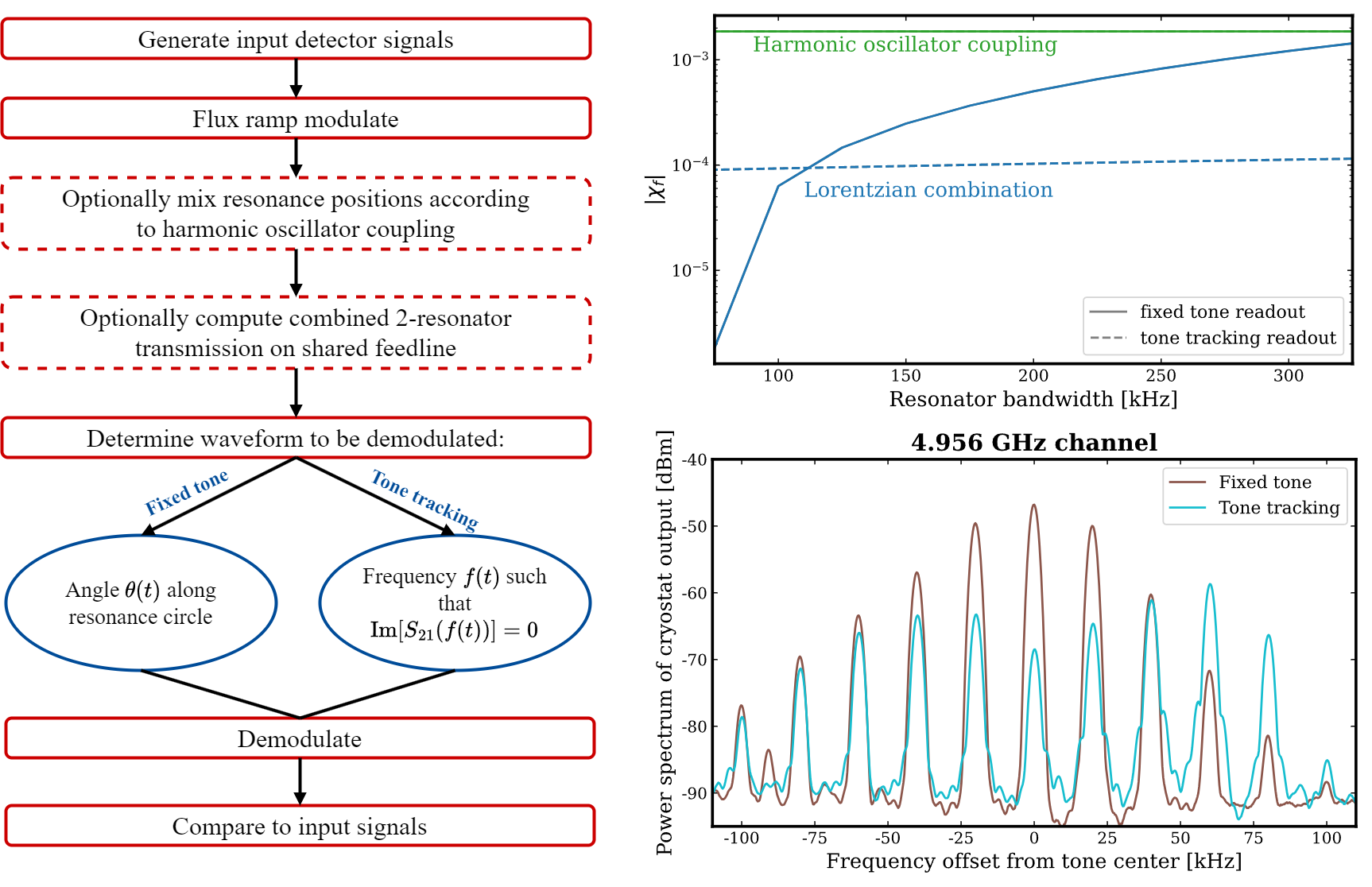}
\caption{(Color figure online) \textit{Left:} schematic of the time-domain simulation used to evaluate the impact of readout scheme on crosstalk.  \textit{Top right:} Simulated $|\chi_f|$ arising from the two frequency crosstalk mechanisms for two channels separated by 1.8 MHz, at a central frequency of 5.0 GHz, with an internal $Q$ of $1.5\times 10^5$, and the SQUID response set to equal the resonator bandwidth.  The Lorentzian contribution is negative, but an absolute value has been taken for a convenient magnitude comparison.  \textit{Bottom right:} Measurement of the amplified transmitted tone and flux ramp modulation sidebands before channelization and demodulation in the vicinity of a single resonance with a 5$\Phi_0$ amplitude 4 kHz flux ramp present but no detector signal.  For this channel, the amplitude of the $j=3, k=0$ feature sets the nonlinear mixing crosstalk level.\label{fig:sim}}
\end{center}
\end{figure}

We evaluate the effect of the two frequency crosstalk mechanisms via a time domain simulation suite which models the effects of flux ramp modulation and demodulation but not those of nonequilibrium resonator physics.  Flux ramp demodulation in the case of fixed tone readout is implemented using the method described in \cite{gard2018fixedtone}.  A schematic illustrating the simulation approach is shown in the left of Figure~\ref{fig:sim} along with example outputs on the upper right.  Unsurprisingly, we find no difference between tone tracked and fixed tone readout on the amplitude of harmonic oscillator coupling crosstalk.  Lorentzian crosstalk, however, is affected by the choice of readout scheme.  Tone tracking and fixed tone readout schemes use different proxies for a channel's true resonance position based on the RF transmission at the probe tone frequency.  These proxies are ignorant of neighboring resonators modifying the RF transmission, so they differ in the amount of cross-coupling they induce in a way that depends on the specific shapes of the two resonances in question.

\par
Only one of the two phase crosstalk mechanisms is affected by tone tracking readout.  Since direct inductive coupling crosstalk is induced by currents at signal (i.e. audio) frequencies before flux ramp modulation, the details of the demodulation method and readout scheme do not affect its amplitude.  Crosstalk from nonlinear mixing, however, will be substantially affected.  As shown for an example channel in the lower right of Figure~\ref{fig:sim}, the amplitude of the largest RF feature with $k = 0$ in the transmitted spectrum is greatly reduced with tone tracking.  The magnitude of this reduction will depend on the resonator parameters, but for typical CMB detector readout channels we measure a reduction of the largest amplitude $k=0$ feature by $\sim$10 dB, directly corresponding to a $\sim$10 dB reduction in nonlinear mixing crosstalk.

%%%%%%%%%%%%%%%%%%%%%%%
% Nonlinear crosstalk %
%%%%%%%%%%%%%%%%%%%%%%%
\section{Nonlinear crosstalk}
\label{sec:nonlinear}

\par
\begin{figure}[htbp]
\begin{center}
\includegraphics[width=\textwidth]{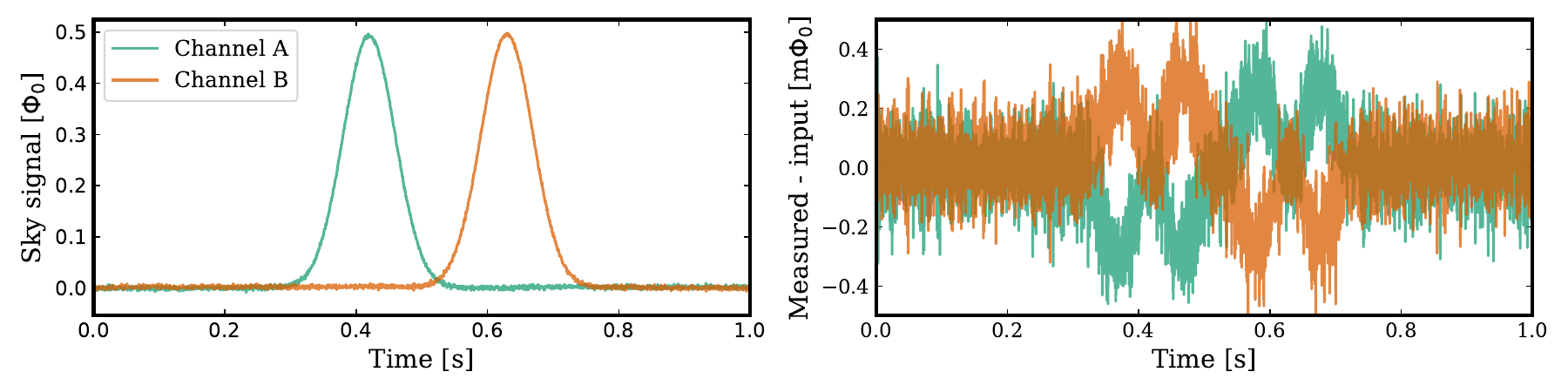}
\caption{(Color figure online) Simulated signal (left) and readout error due to crosstalk (right) from two TES bolometers sensitive to radiation in the 150 GHz atmospheric window in a Simons Observatory Large Aperture Telescope-like instrument scanning across Saturn -- a commonly used calibration source for mm-wave instruments -- using the framework described in Section~\ref{sec:tracking}.  Representative noise, optical efficiency, and atmospheric parameters have been included.  Channel parameters assumed for this simulation are: 1.8 MHz channel spacing, 5.0 GHz center frequency, $1.5\times 10^5$ internal quality factor, 100 kHz bandwidth, and 100 kHz peak-to-peak SQUID response.  Though the signal consists of a single Gaussian peak, multiple peaks are visible in the crosstalk, demonstrating its nonlinear nature. \label{fig:lat_pointsource}}
\end{center}
\end{figure}

The dominant mechanism of crosstalk for the bolometer readout systems discussed in this paper couples resonance frequencies, which are nonlinearly related to the detector signals.  Specifically, for $\chi_f \ll 1$, the signal crosstalk $\chi_\phi$ follows
\begin{equation}
    \chi_\phi \approx \chi_f \cos{(\phi_2 - \phi_1)}
    \label{eq:nonlinear}
\end{equation}
which depends on the flux ramp demodulation phases $\phi_1$ and $\phi_1$ measured in the two channels of interest.  The cosine factor serves to reduce the peak magnitude of the crosstalk at the cost of complicating it.  Figure~\ref{fig:lat_pointsource} shows a simulated example of how this trigonometric relation between frequency and phase crosstalk can complicate the net readout error.

\par
In practice, the phase difference $(\phi_2 - \phi_1)$ will depend on many factors, including the ambient magnetic environment of the multiplexer chip packaging, in-band thermal emission from the enclosing cryostat and intervening atmosphere, telescope pointing and boresight rotation, and observation source intensity.  For a typical daily survey pattern employed by a CMB observatory, the combination of many observations under different weather conditions could partially average down the crosstalk; further study is needed to determine to what extent this may be the case.  Regardless, the nonlinear nature of this crosstalk will significantly complicate any offline time-domain crosstalk removal strategies such as those employed in \cite{henning2015sptpol}.  Finally, by exploiting the elevation angle dependence of in-band atmospheric emission to modulate the $\cos{(\phi_2 - \phi_1)}$ term in Equation~\ref{eq:nonlinear}, future telescopes may be able to measure $\chi_f$ for each channel pair by scanning a sufficiently bright point source at different boresight rotation angles.

%%%%%%%%%%%%%%
% Conclusion %
%%%%%%%%%%%%%%
\section{Conclusion}
\label{sec:conclusion}

\par
Microwave SQUID multiplexed readout utilizes microwave techniques to greatly expand its channel carrying capacity, and is a key enabling technology for several upcoming CMB experiments with more than $10^4$ simultaneously observing TES bolometers.  Through modeling and measurement we have shown that a single physical mechanism of crosstalk is dominant in the NIST multiplexer design relevant for upcoming CMB observatories, with a typical magnitude of $1.5\times 10^{-3}$.  We further find that the impact of tone tracking readout -- as compared to fixed tone readout -- is negligible on the magnitude of crosstalk in currently planned CMB experiments, but conclude that the level of crosstalk in future implementations with a different dominant physical mechanism could be affected by the readout choice.  Finally, we note that the crosstalk depends nonlinearly on many factors, complicating its impact on overall instrument sensitivity.

\backmatter

%%%%%%%%%%%%%%%%%%%%
% Acknowledgements %
%%%%%%%%%%%%%%%%%%%%
\bmhead{Acknowledgments}
JCG was supported in part by a fellowship through the National Research Council Research Apprenticeship Program.  Certain equipment, instruments, or materials are identified in this paper in order to specify the experimental procedure adequately.  Such identification is not intended to imply recommendation or endorsement by NIST, nor is it intended to imply that the materials or equipment identified are necessarily the best available for the purpose.

%%%%%%%%%%%%%%
% References %
%%%%%%%%%%%%%%
\bibliography{Groh_LTD20_umuxcrosstalk}% common bib file

\end{document}